# Estimating Microwave Power Spectra

MAX TEGMARK

*Max-Planck-Institut für Physik*
*Föhringer Ring 6*
*D-80805 München, Germany*
*max@mppmu.mpg.de*

## INTRODUCTION

A new method for estimating the power spectrum $C_\ell$ from cosmic microwave background (CMB) maps was recently presented by the author[1] and applied to the 2 year COBE data, giving the results in Figure 1. It was found that the spectral resolution $\Delta\ell$ for COBE could be more than doubled at $\ell = 15$, thereby revealing previously unresolved features in the power spectrum. Whereas that paper was rather technical, the present paper shows that all qualitative features of this method can be understood from a simple analogy with quantum mechanics.

There has been a surge of interest in the cosmic microwave background radiation (CMB) since the first anisotropies of assumed cosmological origin were detected by the COBE DMR experiment[2]. On the experimental front, scores of new experiments have been carried out and many more are planned or proposed for the near future.

On the theoretical front, considerable progress has been made in understanding how the CMB power spectrum $C_\ell$ depends on various cosmological model parameters, both analytically[3] and quantitatively[4]. It is therefore quite timely to further strengthen the link between these two fronts, by better understanding how to extract more accurate power spectra from experimental data. Early work on this problem[5,6] has recently been extended and applied to the 2 year COBE data[7]. When estimating power spectra, it is customary to place both vertical and horizontal error bars on the data points, as in Figure 1. The former represent the uncertainty due to noise and cosmic variance, and the latter reflect the fact that an estimate of $C_\ell$ inadvertently also receives contributions from other multipole moments. As is well-known, this unavoidable effect is caused by incomplete sky coverage, which destroys the orthogonality of the spherical harmonics. For typical ground- and balloon-based experiments probing degree scales, the spectral blurring $\Delta\ell/\ell$ tends to be of order unity, which makes it difficult to resolve details such as the number of Doppler peaks. By expanding the COBE sky map in spherical harmonics[7], the spectral resolution $\Delta\ell/\ell$ can be brought down to the order of 25%. The optimal method[1] reduces the horizontal error bars still further, down to their theoretical minimum, which is seen to be $\Delta\ell \approx 1$ for the COBE data in Figure 1. It involves solving a generalized eigenvalue problem numerically, were the matrices involved depend on the survey geometry and the pixel variances.

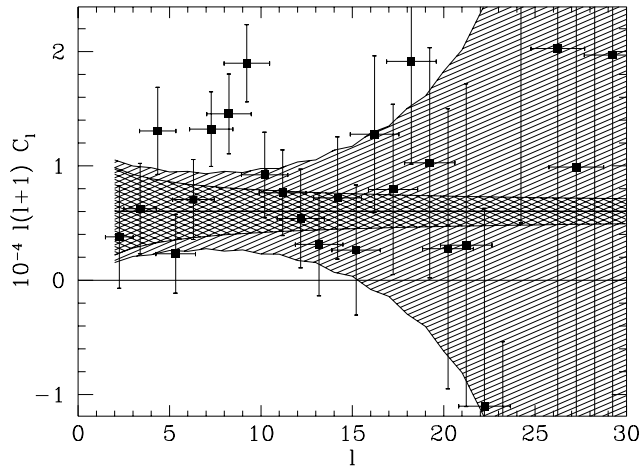

**Figure 1. The power spectrum observed by COBE.**
The observed multipoles are plotted with $1 - \sigma$ vertical error bars, including both pixel noise and cosmic variance. The horizontal error bars show the r.m.s. window function widths. If the true power spectrum is given by $n = 1$ and $Q_{rms,ps} = 20\mu K$ (heavy horizontal line), then the shaded region gives the $1 - \sigma$ error bars and the double-shaded region shows the contribution from cosmic variance.

## THE QUANTUM ANALOGY

Although all the elements of realism included in the numerical technique[1] are important when applying it to real data, one can in fact understand all of the qualitative features of what is going on by ignoring most of these complications.

First of all, let us ignore the fact that a real CMB sky map is pixelized. We let the function $x(\hat{\mathbf{r}})$ denote the temperature fluctuation $\Delta T/T$ in the direction of the unit vector $\hat{\mathbf{r}}$. In the discrete real-world case, each pixel has some r.m.s. noise variance $\sigma^2$ and effectively covers some solid angle $\Omega$ of the sky. For our continuous analogue, let us define the noise variance function $V(\hat{\mathbf{r}})$ by $V \equiv \sigma^2\Omega$, where $\sigma$ and $\Omega$ refer to a pixel in the direction $\hat{\mathbf{r}}$. If we neglect contamination problems, it is easy to see that $V(\hat{\mathbf{r}})$ is simply proportional to the inverse of the time spent observing the direction $\hat{\mathbf{r}}$ (the small-scale details of the pointing of the antenna are irrelevant, as the beam smearing will ensure that the function $V$ is smooth on angular scales below the beam width). Secondly, let us ignore the nuisance terms[8,1] from monopole, dipole, *etc.*, so that we can omit $A$ from equation (1) of [1]. Finally, we let $C_\ell$ denote the power spectrum that results after the experimental beam smearing has been taken into account (this is simply the real power spectrum multiplied by some coefficients that approach zero for very large $\ell$, corresponding to angular scales far below the beam width).

Let us focus on some fixed value $\ell = \ell^*$ for now, say $\ell^* = 17$. The analysis is then repeated from scratch for all other $\ell$-values of interest. Since the power $C_{\ell^*}$ has units of $\mu K^2$, we clearly want to estimate it by some quantity $\tilde{C}_{\ell^*}$ that is quadratic in the data. The simplest estimator of this type is

$$\tilde{C}_{\ell^*} \equiv \left| \int \psi(\hat{\mathbf{r}}) x(\hat{\mathbf{r}}) d\Omega \right|^2, \quad (1)$$

where $\psi$ is some function on the sphere, and the most general estimate is readily shown to be just a weighted average of such estimates[1]. A straightforward calculation shows that

$$\langle \tilde{C}_{\ell^*} \rangle = \sum_{\ell=0}^{\infty} \sum_{m=-\ell}^{\ell} |\hat{\psi}_{lm}|^2 C_\ell + \int |\psi(\hat{\mathbf{r}})|^2 V(\hat{\mathbf{r}}) d\Omega, \quad (2)$$

where $\hat{\psi}_{lm}$ denotes the spherical Fourier transform of $\psi$, *i.e.*, the coefficients in an expansion of $\psi$ in spherical harmonics. In other words, we see that the expectation value of our estimator is the sum of two terms of quite different character. The first, the contribution from cosmology, is the power spectrum convolved with a window function $W_\ell \equiv \sum_m |\widehat{\psi}_{lm}|^2$. The second, the contribution from noise, is just an average value of $V$, the weights being $|\psi(\widehat{\mathbf{r}})|^2$. This is very similar to the result when estimating the power spectrum from a galaxy survey[9]. Just as in that paper, we will find it very convenient to use the standard Dirac quantum mechanics notation with kets, bras and linear operators. This allows us to write $\widehat{\psi}_{\ell m} = \langle \ell m | \psi \rangle$. A window function should always integrate to unity, so the correct normalization for $\psi$ is just $\langle \psi | \psi \rangle = 1$. Defining the operator

$$L \equiv \sum_{\ell=0}^{\infty} \sum_{m=-\ell}^{\ell} \ell |\ell m\rangle \langle \ell m|, \qquad (3)$$

equation (2) becomes simply

$$\langle \tilde{C}_{\ell^*} \rangle = \langle \psi | C_L + V(\widehat{\mathbf{r}}) | \psi \rangle. \qquad (4)$$

Note that $L$ is a scalar operator satisfying $L|\ell m\rangle = \ell |\ell m\rangle$, and is related to the (vector) angular momentum operator $\mathbf{L} = -i\mathbf{r} \times \nabla$ through $\mathbf{L}^2 = L(L+1)$.

Now what is the best choice of $\psi$? Equation (1) tells us that $\tilde{C}_{\ell^*}$ is the square modulus of a random variable whose real and imaginary parts are both Gaussian. Thus if $\psi$ is real, the standard deviation of $\tilde{C}_{\ell^*}$ (the vertical error bar) is simply $\sqrt{2}$ times its expectation value. (If the real and imaginary parts contribute equally, this decreases by a factor of $\sqrt{2}$.) Basically, we minimize the vertical error bars by minimizing $\langle \tilde{C}_{\ell^*} \rangle$. But assuming that our window function is narrow enough that we are measuring mostly what we want to measure, $C_\ell$, the first term in equation (4) satisfies $\langle \psi | C_L | \psi \rangle \approx C_\ell$, independent of $\psi$, so we minimize the vertical error bars by simply minimizing the second term, $\langle \psi | V(\widehat{\mathbf{r}}) | \psi \rangle$. As a measure of the horizontal error bars, we will use $\Delta \ell$, the r.m.s. deviation of the window function from $\ell^*$. With our quantum notation, we have simply $\Delta \ell^2 = \langle \psi | (L - \ell^*)^2 | \psi \rangle$. As we will see, it is impossible to minimize both error bars at the same time, since there is a trade-off between them. It would be like asking for the best and cheapest car. Instead, the best we can do is minimize some linear combination $E \equiv \langle H \rangle$, where we have defined

$$H \equiv (L - \ell^*)^2 + \beta V(\widehat{\mathbf{r}}), \qquad (5)$$

and the parameter $\beta$ specifies how concerned we are about the vertical error bar relative to the horizontal one. Continuing our quantum analogy, we see that we want to find the $\psi$ that minimizes the total "energy", where the "kinetic energy" $(L - \ell^*)^2$ corresponds to the horizontal error bar and the "potential energy" $\beta V(\widehat{\mathbf{r}})$ corresponds to the vertical error bar. If we for sake of illustration set $\ell^* = 1/2$, we simply want to minimize $\langle \psi | \mathbf{L}^2 + \beta V(\widehat{\mathbf{r}}) | \psi \rangle$, given the constraint $\langle \psi | \psi \rangle = 1$. Introducing a Lagrange multiplier $E$, we arrive at the Schrödinger equation

$$[\mathbf{L}^2 + \beta V(\widehat{\mathbf{r}})] | \psi \rangle = E | \psi \rangle. \qquad (6)$$

In other words, we want to find the ground state wavefunction for a particle confined to a sphere with some potential. From our knowledge of quantum mechanics, we can immediately draw a number of conclusions about the solution, all which turn out to agree well with the exact numerical results[1].

- $|\psi|^2$ will be small in regions where the noise variance is large, so regions that received little observation time will receive low weights in the analysis.
- Except for the case of complete sky coverage, we will have $\Delta\ell > 0$.
- If incomplete sky coverage confines $\psi$ to a region of the sky whose angular diameter in the narrowest direction is of order $\Delta\theta$, then the uncertainty principle tells us that the minimum $\Delta\ell$ must be at least of order $1/\Delta\theta$.
- This limit on the spectral resolution is independent of $\ell$.
- For a sky map of COBE type, where $\Delta\theta$ is of order a radian given a 20° galactic cut, the uncertainty principle thus gives $\Delta\ell \gtrsim 1$. This agrees well with the horizontal error bars actually attained in Figure 1.
- If the sky-coverage is incomplete, $V$ is infinite outside of the region covered, and we recover the quantum-mechanical particle-in-a-box problem. From this we know that $\psi$ will always go to zero smoothly as it approaches the survey boundary.
- This smoothness of $\psi$ is really the gist of the method, as it radically reduces "ringing" in Fourier space, "kinetic energy", without increasing the "potential energy" $\langle\psi|V(\hat{\mathbf{r}})|\psi\rangle$ much at all.
- If the the survey volume consists of several disconnected parts, then $\Delta\ell$ is limited by the $\Delta\theta$ of the largest part. For the galaxy-cut COBE case, for instance, using only the northern half of the sky gives the same $\Delta\ell$ as using both the northern and southern skies combined. (However, including both of course helps reduce the *vertical* error bars.)

What typically happens when using the full optimization machinery[1] is that several solutions $\psi$ are found to be almost equally good. For the COBE case, for instance, the ground state solution for a given $\ell^*$ comes out to be almost degenerate, with $(2\ell^* + 1)$ orthogonal eigenfunctions all giving almost the same $E$. This is because the azimuthal symmetry of the galactic cut preserves the orthogonality of spherical harmonics corresponding to different $m$-values. When this happens, we obviously want to average the $(2\ell^* + 1)$ different estimates of $C_{\ell^*}$, as this reduces the vertical error bars by a factor of $\sqrt{2\ell^* + 1}$ without hardly widening the horizontal ones at all. This is basically how Figure 1 was made.

The author wishes to thank Ted Bunn, George Efstathiou, Carlos Frenk and Joseph Silk for useful comments on the manuscript.